\documentclass[floatfix,superscriptaddress,amssymb,twocolumn,aps]{revtex4}

\usepackage{graphicx,dcolumn,bm,amsmath,color}
\usepackage{multirow}
\usepackage{makecell}

\emergencystretch=\maxdimen
\usepackage[version=3]{mhchem} 

\newcommand{\Will}[1]{\textcolor{black}{#1}}

\definecolor{forestgreen}{rgb}{0.13, 0.55, 0.13}

\begin{document}

\title{Molecular Simulations of Quantized Lamellar Thickening in Polyethylenes with Regularly Spaced Brominated Groups}

\author{Kutlwano Gabana}
\affiliation{Department of Physics and Astronomy, University of Sheffield, Sheffield, S3 7RH, United Kingdom}

\author{Gillian A. Gehring}
\affiliation{Department of Physics and Astronomy, University of Sheffield, Sheffield, S3 7RH, United Kingdom}

\author{Hendrik Meyer}
\affiliation{Institut Charles Sadron, Universit\'e de Strasbourg \& CNRS, 23 rue du Loess, 67034 Strasbourg Cedex, France.}

\author{Goran Ungar}
\affiliation{Shaanxi International Research Centre for Soft Materials, School of Material Science and Engineering, Xi’an Jiaotong University, Xi’an 710049, China}
\affiliation{Department of Materials Science and Engineering, University of Sheffield, Sheffield S1 3JD, United Kingdom}

\author{Xiangbing Zeng}
\email{x.zeng@sheffield.ac.uk}
\affiliation{Department of Materials Science and Engineering, University of Sheffield, Sheffield S1 3JD, United Kingdom}

\author{William S. Fall}
\email{william.fall@universite-paris-saclay.fr}
\affiliation{Institut Charles Sadron, Universit\'e de Strasbourg \& CNRS, 23 rue du Loess, 67034 Strasbourg Cedex, France.}
\affiliation{Laboratoire de Physique des Solides - UMR 8502,
CNRS, Universit\'e Paris-Saclay, 91405 Orsay, France}

\begin{abstract}
\Will{Polyethylene (PE) chains, with CH\textsubscript{2} groups replaced by CBr\textsubscript{2} at regular intervals (``precision PE''), have been observed to exhibit competing polymorphs driven by a preference for quantized fold lengths by \citet{tasaki2014polymorphism}. Motivated by this recent discovery, the crystallisation behaviour of such precision PE chains, 400 carbons long with CBr\textsubscript{2} groups placed regularly at every 21st carbon, is investigated using molecular dynamics simulations. The united-monomer model of PE is extended to include \Will{dibromo} groups, with steric clashes at the bromines reflected in a triple-well bending potential, demonstrating its function as a preferred fold site. Different crystallisation protocols, continuous-cooling and self-seeding, reveal remarkably different crystals. Using self-seeding, the crystalline lamellar thickness increases monotonically with temperature, in quantized multiples of the distance between \Will{dibromo} units. Polymer chains are observed to fold preferentially at the dibromo groups and such groups appear to be tolerated within the crystal lamellae. On quenching the bromos assemble to form registered layers, not unlike Smectic phases observed in liquid crystals, which confirms the experimental observation of competing Form I and Form I' polymorphs.}
\end{abstract}

\date{\today}

\maketitle

\section{Introduction}
Most industrial or commodity plastics are semi-crystalline and polydisperse, meaning they are formed from very long chains of different lengths \cite{ungar2001learning} with often a poorly controlled placement of different chemical groups along the backbone \cite{ouchi2018sequence}. It is this inherent dispersity which prevents a clear understanding of the relationship between changes in molecular architecture, such as the substitution of short branches \cite{zeng2007semicrystalline} or other chemical groups along the chains \cite{ouchi2018sequence} \Will{and polymer morphology} and ultimately material properties. Polymer crystallisation is a kinetically driven, non-equilibrium process and whilst thermodynamics can predict which structure is preferred at a given temperature or pressure, in thermal equilibrium, it is kinetics \Will{which determines the fastest-growing morphology and} crystal form, as well as the organisation of chains within the semi-crystalline morphology \cite{cheng2005enthalpic}. To \Will{circumvent the problems of polydispersity and non-uniformity introduced by polymerization}, many choose to work with narrow molecular weight distributions or monodisperse systems where competing theories of polymer crystallisation can be tested with precision. 

Polyethylene is the simplest homopolymer and unsurprisingly one of the most widely studied. Many experimental \cite{ungar2001learning,ungar1979long,buckley1984structure,kovacs1975isothermal} and theoretical studies \cite{gee2006atomistic} have been undertaken on monodisperse systems of short PE chains, \Will{i.e.} n-alkanes. \Will{Alkanes longer than 100 carbons were found to be able to crystallize as both extended and once-folded chains, while the longest one synthesized, n-C\textsubscript{390}H\textsubscript{782}, could even be folded in five \cite{ungar2001learning}. Solution-grown crystals have sharp regular folds \cite{ungar1987infra}. In contrast, the initial form grown from the melt has a disordered intercrystalline layer resembling that found in polydisperse polymers \cite{ungar2000non}, a structure well known to be reproduced by simulation \cite{ramos2018review}. } The addition of single branches, placed symmetrically or asymmetrically is noted to induce different nanostructures \cite{zeng2007semicrystalline}, depending on the crystallisation temperature. Remarkable achievements from studies of monodisperse n-alkanes include the discovery of chain folding, influence of chain branching, an improved understanding of the relationship between chain length, crystal thickness and melting temperature, the discovery of self-poisoning \Will{crystallization} \cite{ungar2005effect} and even quasi-continuous melting behaviour in adsorbed monolayers\cite{zhang2021quasi,zhang2021roughening}. \Will{However, the price for obtaining strictly monodisperse polymers is high, as they are obtained not by polymerization but by a sequence of protection-doubling-deprotection-purification steps, the yield suffering exponential decay along the line. Regular placement of chemical groups in a polydisperse polymer is no mean task either. 
However recent advances in synthetic chemistry facilitate such efforts \cite{barner2012new} and new types of periodically substituted polyethylenes have been obtained. Example substituents include}, halogens \cite{zhang2018effect}, acetals \cite{zhang2019crystallization,zhang2020crystallization,marxsen2021crystallization,liu2022effect}, \Will{diesters \cite{marxsen2022crystallization}} and many others which may be used to create hierarchically nanostructured materials \cite{zhang2023chain}. 

It is however, experimentally challenging to pin down the \Will{molecular organization in semicrystalline polymers, including  } precision PE materials due to their inherent semi-crystalline nature. Structures are often inferred as opposed to directly observed and must be supported by additional measurements. Over the past few decades, computer simulations have emerged as a new route to understand the large-scale structures formed by long polymer chains \cite{gartner2019modeling,TangEtal:Macro2019}. This has been enabled by advances at various levels, improved polymer models, highly efficient molecular simulation codes \cite{thompson2022lammps}, technological increases of computational power, and the availability of high-performance computers. 

Simulations of linear PE systems are now commonplace, with all-atom models addressing small-scale structures formed by short chains, which successfully reproduce the orthorhombic crystalline unit cell \cite{ramos2018review}. However large-scale all-atom simulations of PE are near impossible primarily due to the number of atoms required and intra-chain potentials which must be evolved. In addition, where simulations of polymer crystallisation are concerned, nucleating crystals from the melt in long chain systems is very challenging and requires exceptionally long run times. Increasingly, coarse-grained \Will{(CG)} models are being used to overcome these computational bottlenecks. United-atom (UA) models, which absorb hydrogen atoms into the heavy atom they are connected to, offer a more simplistic view of the chain backbone, due to the reduction in intra-chain angular and torsional potentials. However this comes at the expense of not being able to reproduce the correct unit cell of crystalline PE \cite{zhang2018direct}. Despite this, UA models have been used extensively to study the crystalline properties of PE systems \cite{ramos2018review,kumar2017effect,waheed2002molecular,yi2013molecular}. Addressing large-scale structures typical of PE crystals, i.e. semi-crystalline lamella upwards of 10nm, requires heavier coarse-graining. United-monomer models, first introduced some 20 years ago for poly-vinyl alcohol \cite{meyer2001formation,meyer2002formation,reith2001mapping,luo2013disentanglement,luo2016role,jabbari2015plastic,jabbari2018static} (PVA), have shown recent success when adapted for PE \cite{fall2022role,fall2023molecular}. Such models capture the distinct torsional states of PE with characteristic angular potentials extracted from all-atom simulations using Iterative Boltzmann Inversion (IBI) \cite{reith2001mapping,mooreIBI}. Other simpler CG models have also been used \cite{hall2019coarse,hall2019divining,zhai2019crystallization} to study PE crystallisation. In the study of melt properties it is common to employ CG strategies with even higher levels of coarse-graining \cite{GeRoFa2004jcp,SaAgPeGr2016prl,SaBe2018jctc}\Will{. However} where PE crystallisation is concerned a CG unit representing more than two carbons is too far removed from the real polymer. This results in many important chain conformations being coarse-grained-out and makes it challenging to relate results to the real polymer. Simulations of precision PE systems with controlled branching are however rather rare and those with regularly placed chemical moieties are virtually non-existent. 

In recent studies, united monomer models have demonstrated control over the lamellar thickness in regularly branched PE by growing crystals using a technique known as self-seeding \cite{fall2022role,fall2023molecular}, \Will{first introduced experimentally in the 1960s by Blundell and Keller \cite{blundell1968nature}}. Motivated by this recent success, an extension of the united monomer model of PE is developed here to include \Will{di-halogen units}, in this case \Will{dibromo}, in order to study the effect of regularly placed chemical groups on the large-scale semi-crystalline morphology. The crystallisation behaviour of monodisperse systems of precision PE chains, with \Will{dibromomethylene groups}, is investigated using two different crystallization protocols, continuous-cooling and self-seeding \Will{isothermal crystallization}, for a series of different cooling rates and crystallisation temperatures. Finally we conclude and compare our results to recent experimental studies in PEBr systems \cite{zhang2018effect,tasaki2014polymorphism,zhang2020crystallization,marxsen2021crystallization}, noting some important trends and highlighting future questions to be addressed in a forthcoming study in much larger systems. \Will{This is the first molecular dynamics simulation study to demonstrate isothermal crystallisation temperature driven quantized lamellar thickening, in semi-crystalline systems of PE, with regularly placed chemical substituents.}


\section{Model \& Methods}

\begin{figure}
    \centering
    \includegraphics[width=0.95\columnwidth]{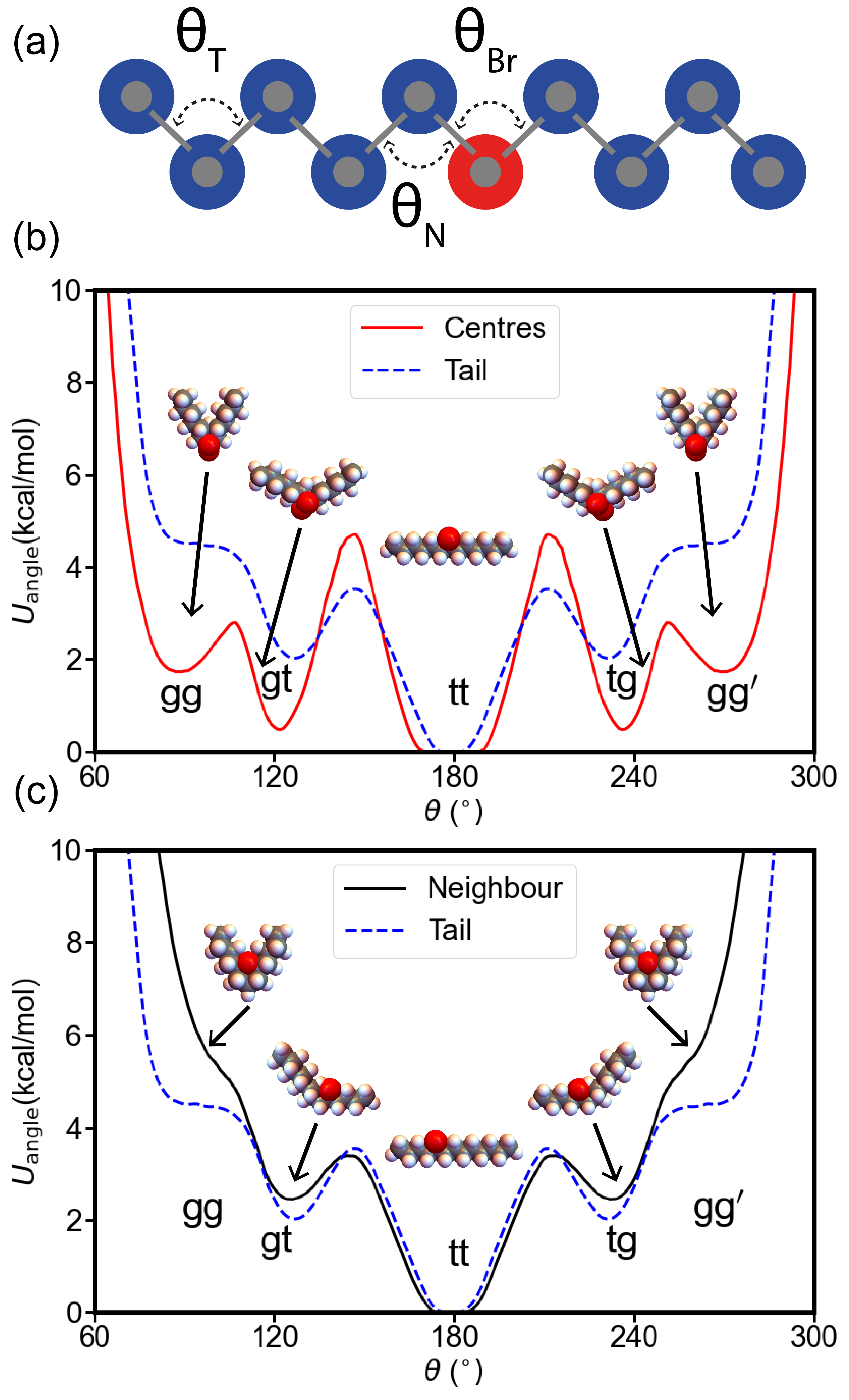}
    \caption{Coarse-grained model and angular potentials. (a) Schematic representation of the coarse-grained model for PE, blue and red beads represent C\textsubscript{2}H\textsubscript{4} and CH\textsubscript{2}CBr\textsubscript{2} CG centres respectively. Different bending potentials considered between bromine centres, their immediate neighbours and tails, namely $\theta_{\mathrm{Br}}$, $\theta_{\mathrm{N}}$ and $\theta_{\mathrm{T}}$ are drawn in. (b-c) Mapping from an all-atom torsional potential to CG tabulated angular potential, for the bromine centres ($U(\theta_{\mathrm{Br}})$, red curve) and their neighbours ($U(\theta_{\mathrm{N}}$), black curve) respectively shown alongside the potential of the tail groups. Note the angular potential of the tails (U($\theta_{\mathrm{T}}$), dashed blue curve) is identical to that of regular polyethylene reported previously \cite{fall2022role}. The minima of the CG potentials map to real all-atom conformers of precision PE chains as depicted in the inset artistic representations at each of the respective minima.}
    \label{fig:angular_potentials}
\end{figure}

An extension of the united-monomer model, well established to capture the behaviour of a variety of macromolecules \cite{meyer2001formation,Mey2003lnp,fall2022role,fall2023molecular}, is presented here for precision polyethylenes, \Will{ i.e. $n$-alkanes, with CBr\textsubscript{2}} groups placed \Will{periodically} along the chains. In the united-monomer model, chemical monomers C\textsubscript{2}H\textsubscript{4} and CH\textsubscript{2}CBr\textsubscript{2} are represented by single coarse-grained beads as denoted by the blue and red species in Fig. \ref{fig:angular_potentials} (a). The CG procedure employed uses every second carbon along the all-atom chain as the mapping point (or CG center) as opposed to the centre-of-mass, known to result in cross-correlations between angles and bonds for low CG levels \cite{reith2001mapping}, preserving the exact mapping of the angular potential between all-atom and CG backbones. Stretching valence terms take the form of harmonic potentials, see Table \ref{tbl:params},
where $U_{\mathrm{bond}}$ represents the change in potential energy associated with deforming the bond away from its equilibrium separation, $l_{0}$. This is identical to the united-monomer model of PE previously reported \cite{fall2022role} and remains unchanged for \Will{dibromo} monomers CH\textsubscript{2}CBr\textsubscript{2}. Details of the all-atom simulations and Boltzmann Inversion procedure used to extract effective potentials can be found in Supporting Information. 

Bending valence terms, in the original united-monomer model of PE, were obtained via Boltzmann Inversion of the bond angle distribution, $P(\theta)$ between CG centers in all-atomistic simulations. In this procedure the bond angle probability distribution $P(\theta)$, is inverted to produce a unique bending potential $U_{\mathrm{angle}}(\theta)=-k_{\mathrm{B}}T\ln{[P(\theta)}/\sin\theta]$. Thus the torsional angles in the all-atomistic model become effective angular potentials between every three successive CG beads, resulting in the tabulated potential shown by the dashed (blue) line in Figs. 1 (b) and (c). Including additional \Will{dibromo} groups required consideration of two additional CG centers, one at the \Will{dibromo} group (Centers) and another next to the \Will{dibromo} group (Neighbours) as well as the standard PE chain (Tails), which is already known from the united monomer model of PE \cite{fall2022role}. We note that further CG centers were not required since the bond angle distribution rapidly collapses onto standard PE, when sufficiently far away from the Br group. The angular potentials next to the Br dimer retain the characteristic triple-well shape of the standard PE model but with a more strongly disfavoured \textit{gauche-gauche} state, indicated by a more prominent shoulder at shallow angles in the (black) curve in Fig \ref{fig:angular_potentials} (c). The key difference arises at the Br group, which instead resembles a quintuple-well potential, see the (red) curve in Fig. \ref{fig:angular_potentials} (b), with an additional deep \Will{minimum} present at the \textit{gauche-gauche} state and a higher torsional barrier between \textit{trans-gauche} and \textit{trans-trans} states. Thus chain-folding or bending is more strongly favoured at the Br group and more strongly disfavoured at nearest-neighbour groups. This feature is not unexpected, with recent experiments reporting \Will{chevron-like} crystals, with integer numbers of Br atoms between successive chevrons, suggesting a strong tendency to bend at CBr\textsubscript{2} monomers \cite{zhang2018effect}. This behaviour is further confirmed on examination of the different conformers along the all-atomistic chain, see the inset cartoons in Figs. \ref{fig:angular_potentials} (b) and (c). It is noteworthy that \textit{gg} and \textit{gt} conformers taking place at the \Will{dibromo group} move neighbouring H atoms along the chain further away from the rather bulky Br group. On the other hand, when the Br group is located on the neighbouring monomer, H and Br atoms appear to clash and are brought closer in a \textit{gg} conformer hence the pronounced shoulder at shallow angles.

\begin{table*}
\caption{\label{tbl:params} Potential forms and parameter values for the united-monomer model of PE with \Will{dibromo} groups in real units. }
\begin{tabular}{c|c|c|c}
& Interaction & Form & Parameters \\
\hline
\hline
\multirow{3}{*}{Bonds} & \multirow{3}{*}{\makecell{(C\textsubscript{2}H\textsubscript{4})-(C\textsubscript{2}H\textsubscript{4})\\
(CH\textsubscript{2}CBr\textsubscript{2})-(C\textsubscript{2}H\textsubscript{4})}} & \multirow{3}{*}{$U_{\mathrm{bond}}(l)\texttt{=}\frac{1}{2}k_{\mathrm{bond}}(l-l_{0})^{2}$} & \multirow{3}{*}{\makecell{$k_{\mathrm{bond}}\texttt{=}53.69$ (kcal/mol/\AA\textsuperscript{2})\\ $l_{0}\texttt{=}2.225$ (\AA)}}  \\
& & & \\
& & & \\
\hline
\multirow{3}{*}{Angles} & (C\textsubscript{2}H\textsubscript{4})-(C\textsubscript{2}H\textsubscript{4})-(C\textsubscript{2}H\textsubscript{4}) & Triple-Well Tab &  Fig.1(b,c)-Tails (blue) \\
\cline{2-4}
& (C\textsubscript{2}H\textsubscript{4})-(C\textsubscript{2}H\textsubscript{4})-(CH\textsubscript{2}CBr\textsubscript{2}) & Triple-Well Tab & Fig.1(c)-Neighbors (black)\\
\cline{2-4}
& (C\textsubscript{2}H\textsubscript{4})-(CH\textsubscript{2}CBr\textsubscript{2})-(C\textsubscript{2}H\textsubscript{4}) & \Will{Triple}-Well Tab & Fig.1(b)-Centres (red)\\
\hline
\multirow{7}{*}{\makecell{Non-Bonded}} & (C\textsubscript{2}H\textsubscript{4})$\leftrightarrow$(C\textsubscript{2}H\textsubscript{4}) & \multirow{4}{*}{\makecell{\(\scalebox{1.2}{$\,U_{\mathrm{LJ}}^{({9\texttt{-}6})}\texttt{=}4\epsilon_{0}\bigg[\Big(\frac{\sigma_{0}}{r}\Big)^{9}\texttt{-}\Big(\frac{\sigma_{0}}{r}\Big)^{6}\bigg]$}\)}} & \makecell{$\epsilon_{0}\texttt{=}0.348$ (kcal/mol)\\$\sigma_{0}\texttt{=}4.45$ (\AA) \\ $r_{\mathrm{c}}=(3/2)^{\frac{1}{3}}\sigma_{0}$}   \\
\cline{2-2} 
\cline{4-4}
& (CH\textsubscript{2}CBr\textsubscript{2})$\leftrightarrow$(C\textsubscript{2}H\textsubscript{4}) & & \makecell{$\epsilon_{0}\texttt{=}0.44$ (kcal/mol)\\$\sigma_{0}\texttt{=}5.19$ (\AA) \\ $r_{\mathrm{c}}=1.5^{\frac{1}{3}}\sigma_{0}$} \\
\cline{2-4}
& (CH\textsubscript{2}CBr\textsubscript{2})$\leftrightarrow$(CH\textsubscript{2}CBr\textsubscript{2}) & \(\scalebox{1.2}{$U_{\mathrm{LJ}}^{({12\texttt{-}6})}\texttt{=}4\epsilon_{0}\bigg[\Big(\frac{\sigma_{0}}{r}\Big)^{12}\texttt{-}\Big(\frac{\sigma_{0}}{r}\Big)^{6}\bigg]$}\) & \makecell{$\epsilon_{0}\texttt{=}0.45$ (kcal/mol)\\$\sigma_{0}\texttt{=}5.66$ (\AA) \\ $r_{\mathrm{c}}=2^{\frac{1}{6}}\sigma_{0}$} \\
\hline
\end{tabular}%
\end{table*} 

Non-bonded soft Van der Waals interactions between CG beads, separated by three or more successive bonds, are modelled in a hybrid fashion using either 9-6 or 12-6 Lennard-Jones (LJ) potentials, see Table.\ref{tbl:params}. Both the standard PE (tails), in keeping with the previous model \cite{fall2022role} and mixed interactions between Br (centre) and PE (tail) groups are described \Will{by} soft pairwise 9-6 LJ potentials 
where $\epsilon_{0}$ denotes the depth of the potential well, $\sigma_{0}$ the zero-crossing (particle size) as defined in Table.~\ref{tbl:params}, $r$ the inter-particle separation and $r_{\mathrm{c}}$ the cutoff distance. Both interactions are broadly similar, with mixed interactions between Br groups and standard PE groups requiring a marginally deeper minima which is shifted to larger separations to accommodate the larger Br atoms. A 9-6 LJ potential is chosen to reflect the softer effective interactions between united-monomers. Two Br groups however interact by a less-soft 12-6 LJ potential which 
provides a marginally better approximation to the more spherical CH\textsubscript{2}CBr\textsubscript{2} CG units as opposed to C\textsubscript{2}H\textsubscript{4}  with a much deeper minima and larger separation. Both potentials are cut and shifted to zero at the their respective cutoffs $r_{\mathrm{c}}$ as defined in Table~\ref{tbl:params}, which ensures only repulsive interactions between CG beads. Crystallisation studies, which include only the repulsive part of the potential, have been demonstrated very successfully for quiescent systems \cite{luo2013disentanglement,luo2016role,meyer2001formation,meyer2002formation,fall2022role} with recent studies reporting the largest multi-lamella polymer crystals grown to date in simulations of PE \cite{fall2023molecular}. We therefore follow this well-established approach. Note electrostatics are not explicitly taken into account in this model as a monomer has no net charge. In addition, more precise potentials could be obtained at specific state points by iterative Boltzmann inversion using tabulated potentials but are not considered here. \cite{RePuMP2003jcc,SaAgPeGr2016prl,moyassari2019molecular}

Simulations consist of 96 chains of length 200 united monomers, with \Will{dibromo} groups placed on every 10th or 11th united monomer unit along the chain. This was chosen to preserve a direct mapping from the CG chain to all-atomistic conformations of PE21, as reported in \cite{zhang2018effect}, in preparation for future studies where atomistic monomers may be reinserted into the large-scale structures obtained here. Initial topologies were taken from simulations of polyethylene melts studied in \cite{fall2022role} and hence were already well equilibrated. \Will{Dibromo} units are then inserted regularly along the linear backbone. As demonstrated previously for small Butyl branches, insertion of a few small units causes a small disturbance of the chains \cite{fall2022role}. All equilibration and production runs use the Large-Scale Atomic/Molecular Massively Parallel Simulator (LAMMPS) \cite{plimpton1995fast,thompson2022lammps}. After insertion of the \Will{dibromo} monomers along the existing PE melt, a short equilibration run is performed to allow the chains to fully relax in the NPT ensemble via a Langevin thermostat, with coupling constant $\mathit{\Gamma}=0.5$ ($1/\tau$) and a Berendsen barostat with $P_{\mathrm{damp}}=100.0$ ($\tau$). The integration timestep used is 0.005~$\tau$, where the LJ-time unit $\tau = \sqrt{m\sigma^2/k_\mathrm{B}T_{0}}$ corresponds to 2.7~ps ($m=27.3881$~g/mol) and temperature is held fixed at $T_0=500$~K $= 227^\circ$C ($T=1.0$ in reduced units). A high pressure, $P=8.0$ ($k_\mathrm{B} T_0/\sigma^3$) is used to obtain the correct density. For production runs a Nos\'e-Hoover thermostat and barostat is used during cooling and self-seeding runs with $T_{\mathrm{damp}}=2.0$ ($\tau$) and $P_{\mathrm{damp}}=100.0$ ($\tau$) respectively.

\begin{figure*}
    \centering
    \includegraphics[width=1.75\columnwidth]{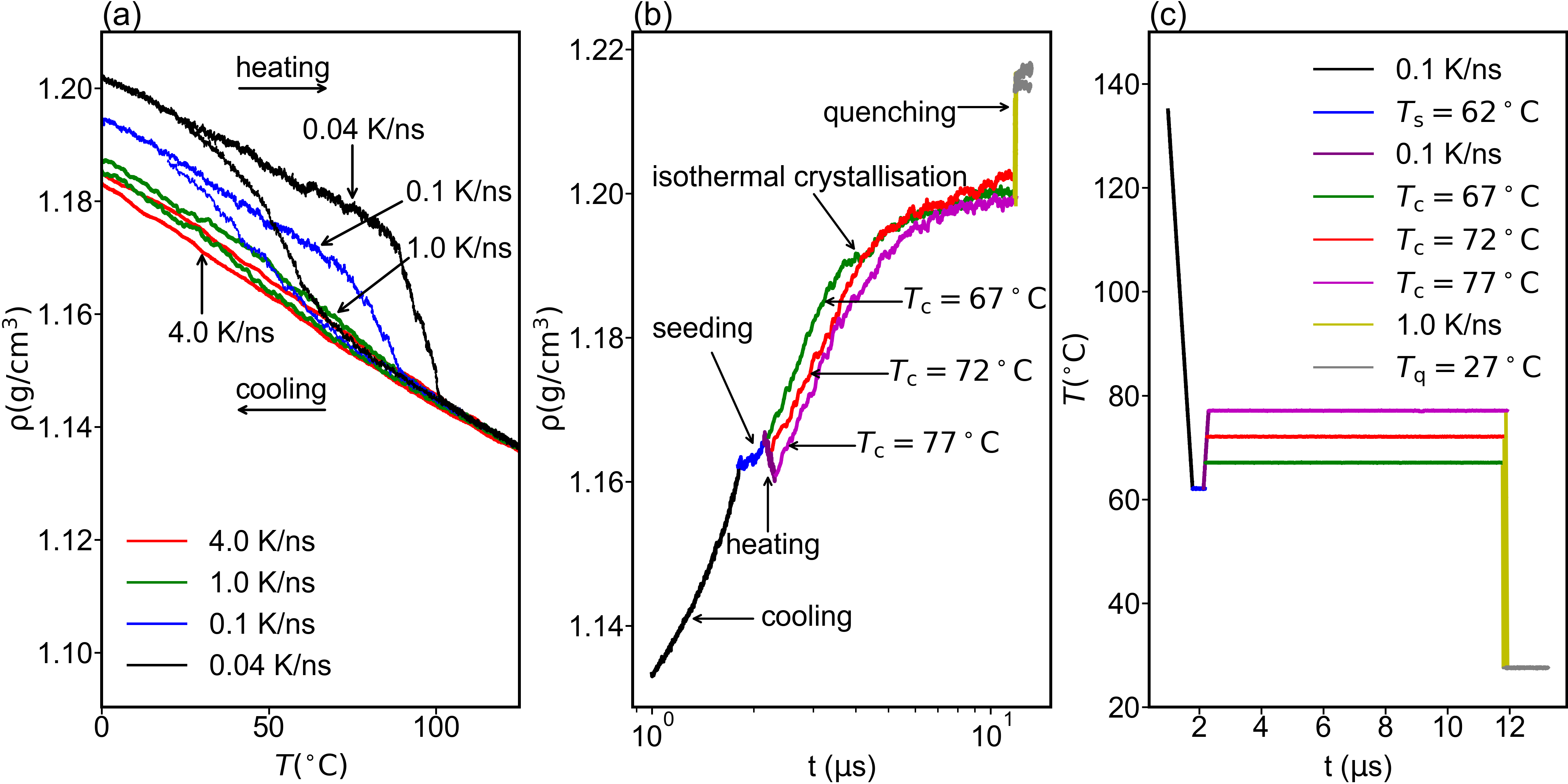}
    \caption{Density curves for all systems during the self-seeding and continuous-cooling simulations. (a) Density-temperature curves for the continuous-cooling starting from 125$^{\circ}$C to 0$^{\circ}$C with four different rates and their corresponding heating curves. The 0.1 K/ns and 0.04 K/ns cooling were bifurcated from the 1.0 K/ns cooling at 127 $^{\circ}$C. (b) Density-time curves during the self-seeding procedure. Each respective cooling, seeding, heating, crystallisation or quenching stage is indicated in the figure with arrows. The seeding (blue) curve is bifurcated from the 0.1 K/ns cooling (black) curve at $T_{\mathrm{s}}=62$ $^{\circ}$C. This is followed by a 0.1 K/ns heating (purple curve) to a series of crystallisation temperatures $T_{\mathrm{c}}=$ 67$^{\circ}$C, 72$^{\circ}$C and 77$^{\circ}$C (green, red and pink curves) respectively at which the systems are held for 0.4$\mu$s and then quenched (yellow curve) to 27$^{\circ}$C at 1.0 K/ns. Note the 0.1 K/ns cooling curve begins at $\sim1 \mu$s due to bifurcation from the faster 1.0 K/ns cooling at 127 $^{\circ}$C for computational expedience. The corresponding time-temperature protocol is similarly shown in panel (c).}
    \label{fig:dens_curve}
\end{figure*}

Two methods of crystallisation were employed in this study to generate crystal structures, namely, continuous-cooling and self-seeding. The continuous-cooling protocol entails a linear ramping of temperature with time from the melt at $T=227^\circ$C to $T=0^\circ$C. In this instance the polymer melt, equilibrated at $227^{\circ}$C, was first cooled at 4.0 K/ns and 1.0 K/ns respectively to room temperature. Two slower cooling rates of 0.1 K/ns and 0.04 K/ns were also then employed but bifurcated from the faster (1.0 K/ns) cooling curve at 127$^{\circ}$C, well above the onset of crystallisation. Continuous-cooling on simulation timescales ($\sim\mu s$) is suitable for the crystallisation of short oligomers. However where long chains are concerned, it is often too fast and results in rapid nucleation and several disordered amorphous regions, making for a poor crystal structure. The second crystallisation method, self-seeding, is used to combat this issue. During self-seeding, the crystallisation process is initiated by the presence of existing crystalline structures within the polymer.

The time-temperature protocol for self-seeding used in this work is illustrated in Fig \ref{fig:dens_curve} (c) and begins similarly by continuously cooling the system at 0.1 K/ns (black curve) to a seeding temperature ($T_{\mathrm{s}}=62^\circ$C). This is then followed by an isothermal run (blue curve), for a few hundred nanoseconds, until the chains begin to crystallise. Note initially the system was allowed to undergo isothermal crystallisation at a range of different temperatures, by bifurcating from the 0.1 K/ns cooling curve. The best crystallisation was found in the system crystallised at $62^{\circ}$C and was therefore chosen as the seeding temperature. After sufficient nuclei have formed, the system is continuously heated at 0.1 K/ns (purple curve) to a series of crystallisation temperatures $T_{\mathrm{c}}$, until most nuclei have melted away. At this point another isothermal crystallisation is carried out, for several microseconds, until the density reaches a plateau indicating the remaining nuclei have grown sufficiently large to fill the box. All systems as grown at their respective crystallisation temperatures $T_{\mathrm{c}}$, were then quenched to room temperature $T_{\mathrm{q}}=27^\circ$C at 1.0 K/ns to gather statistics. By self-seeding, remarkably clean crystals are grown facilitating a direct comparison with recent experimental observations of competing polymorphs in \Will{periodically spaced} PEBr polymers \cite{tasaki2014polymorphism,zhang2018effect}.

\section{Results \& Discussion}

Figure \ref{fig:dens_curve} (a) shows the density-temperature profiles during continuous-cooling for all cooling rates considered, covering two decades. The fastest 4.0 K/ns cooling rate (red curve) shows weak hysteresis behaviour suggesting the absence of crystalline regions and that the structure is almost entirely amorphous. This is contrary to simulations of linear PE chains \cite{fall2022role,meyer2002formation}, where such rapid cooling is known to induce at least some crystallisation, suggesting PE chains with Br units placed regularly along the chains are inherently more challenging to nucleate. With slower cooling at 1.0 K/ns (green curve) some weak hysteresis behaviour is present consistent with poor crystallisation suggesting a lack of well-defined crystalline regions, i.e. the polymer chains have not organized into a highly ordered structure. It is only with the slowest cooling rates reasonably achievable on simulation timescales, at 0.1 K/ns and 0.04 K/ns, that pronounced hysteresis behaviour becomes noticeable in the blue and black curves in Fig \ref{fig:dens_curve} (a). The density can be seen to monotonically increase with slower cooling, consistent with other simulation studies \cite{fall2022role} however, the crystals formed by continuous-cooling are still remarkably poor. To overcome this challenge and grow PEBr crystals representative of real experiments, crystallisation by self-seeding is employed here.

Using the self-seeding procedure outlined in the previous section, an isothermal crystallisation is carried out for upwards of 4 $\mu$s at 3 different temperatures, 67$^{\circ}$C, 72$^{\circ}$C and 77$^{\circ}$C, as shown by the green, red and purple curves respectively in Figure \ref{fig:dens_curve} (b). The crystals formed by continuous-cooling are very poor when compared with those grown via self-seeding. The density and crystallinity at the crystallisation temperature are comparable with that of the continuously-cooled systems at room temperature. The crystallisation also appears faster as seen by the sudden change in the slopes of the density curves. In addition, the density then exceeds that of the continuously cooled systems upon quenching to room temperature, which is already indicative of improved crystallisation.

\begin{figure*}
    \centering
    \includegraphics[width=1.5\columnwidth]{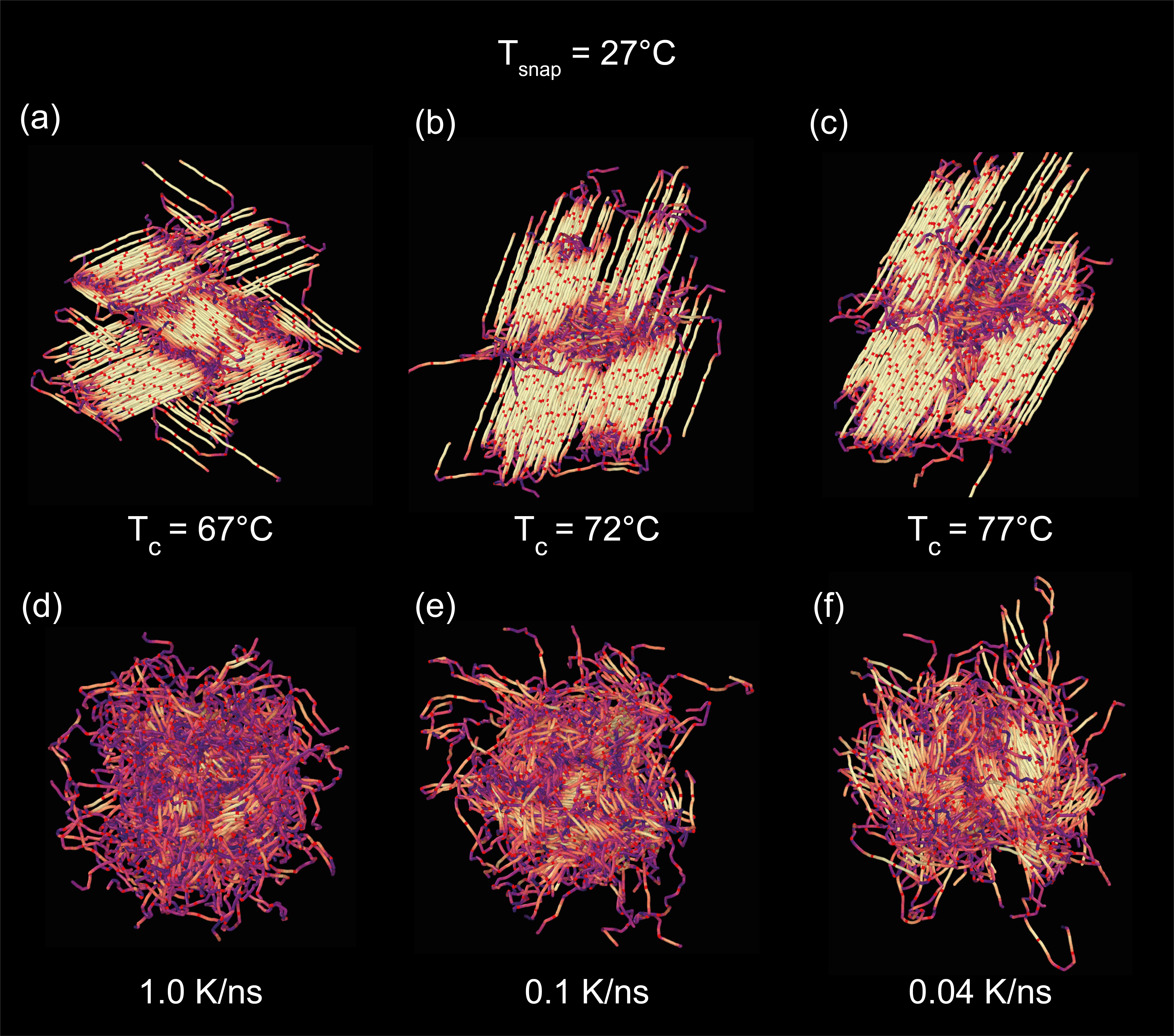}
    \caption{Snapshots of the continuously cooled and self-seeded systems at 27$^{\circ}C$. Panels (a-c) show the self-seeded crystals, grown at 67$^{\circ}$C, 72$^{\circ}$C and 77$^{\circ}$C respectively, after quenching. Panels (d-f) show the crystals grown by continuous-cooling. Polymer chains are coloured continuously according to their local $P_{2}$ order parameter from $P_{2}=0$ (purple, amorphous) to $P_{2}=1$ (yellow, crystalline) and the bromine species is shown in red.}
    \label{fig:Snaps}
\end{figure*}

The difference between the two protocols becomes more evident when examining the snapshots at the end of the cooling runs which are shown in Figure \ref{fig:Snaps} (d-f). Here the polymer chains are coloured continuously according to their local $P_{2}$ order parameter, which may be defined mathematically as the second Legendre polynomial of the cosine of the angle between the orientation of a particle and a reference axis,
\begin{equation}
P_{2}= \frac{\langle 3\cos^2{\theta}-1\rangle}{2} \
\label{e:p2}
\end{equation}
where ⟨⟩ denotes an average over all particles in a local region. Chains are coloured from $P_{2}=0$ (purple, amorphous) to $P_{2}=1$ (yellow, crystalline) and the bromine species is shown in red. For details of the analysis method used to assign crystallinity to beads see \cite{fall2023molecular}. In the case of continuous-cooling, the systems are plagued by large regions of disordered chains, especially for the fastest cooling rates, as shown in the snapshot in panel (d) after cooling at 1.0 K/ns. These structures likely arise due multiple nuclei that grow in different orientations and crash into each other, stopping each other from growing. Moreover the timescale over which the systems are cooled is so short that nucleation events are suppressed to lower temperatures, where the polymer chains are less mobile, resulting in very poor crystallisation. In the system cooled at 1.0 K/ns it appears as though cooling was sufficiently rapid to suppress nucleation almost entirely. The overall crystallinity improves with slower cooling as seen in the 0.1 K/ns and 0.04 K/ns snapshots in panels (e) and (f) respectively, which is not unexpected. The slowest 0.04 K/ns cooling achieved the best crystallisation and has regions where chains are aligned but still resembles a highly amorphous structure. 

\begin{figure*}
    \centering
    \includegraphics[width=1.5\columnwidth]{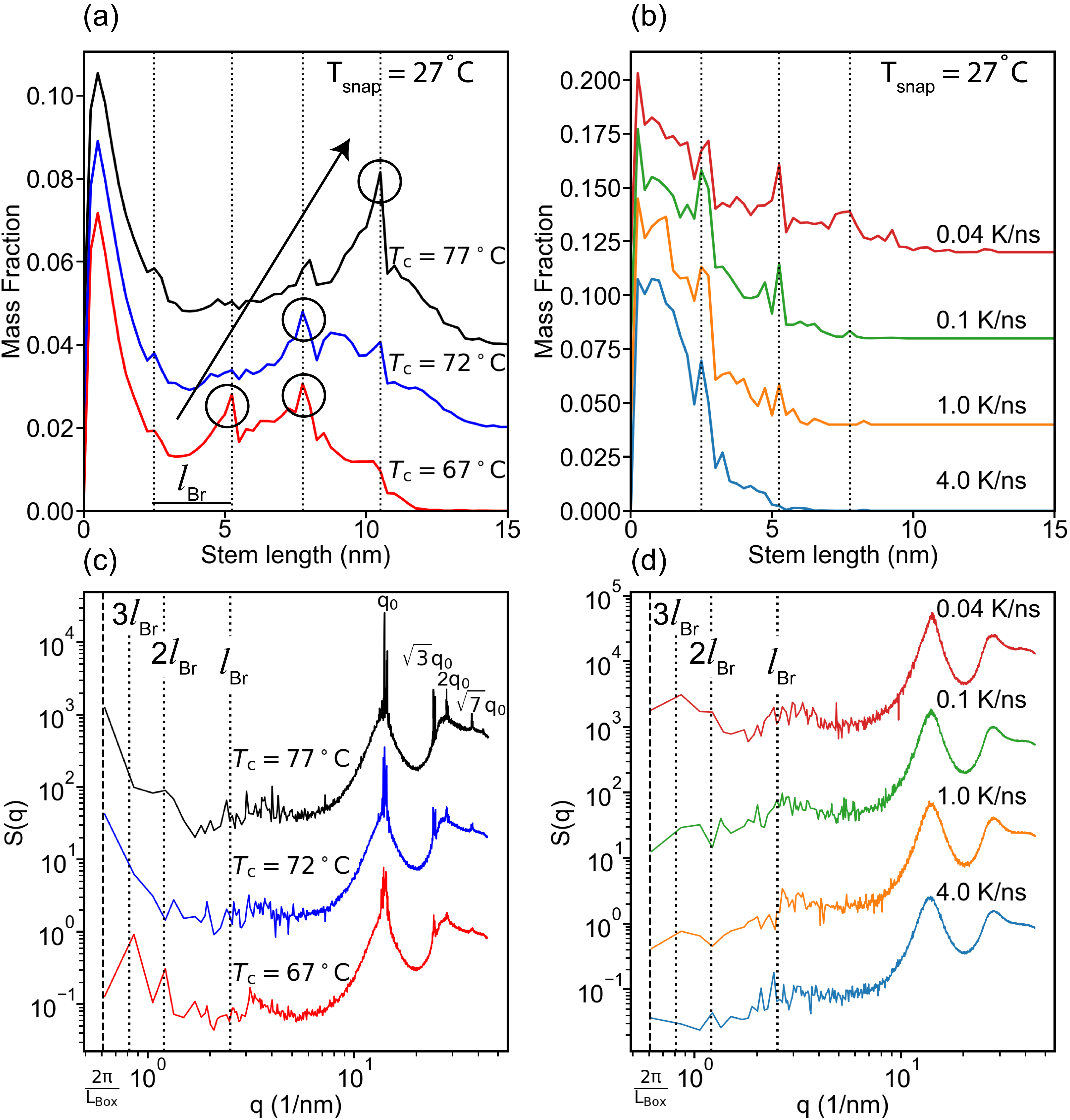}
    \caption{Stem length distributions and static structure factors for all systems at the end of the continuous-cooling and for the self-seeding runs at 27$^{\circ}$C. Panels (a) and (b) show stem length distributions of all systems at the end of the self-seeding and continuous-cooling protocols respectively. The dotted lines represent the distance between neighbouring bromines $l_{Br}$, along the chains. Panels (c) and (d) show the static structure factors after self-seeding and continuous-cooling at 27 $^{\circ}$C respectively. The dashed line represents the limit of the simulation box corresponding to 2$\pi/L_{box}$, dotted lines indicate the distance between neighbouring bromines $l_{Br}$ in the SAXS region of the structure factor and $q_{0}$ the hexagonal lattice parameter.}
    \label{fig:structure}
\end{figure*}

This is in stark contrast to the snapshots from the self-seeding runs in panels (a-c), where clear lamellar structures can be seen with large folded crystalline stems appearing, with well defined amorphous regions at the fringes. For the systems crystallised at 72$^{\circ}$C and 77$^{\circ}$C, panels (b) and (c) respectively, the crystal nearly consumes the whole box and the crystalline chain stems align globally with the same orientation. However, in panel (a), at 67$^{\circ}$C it appears as though two large crystal regions have grown simultaneously and perpendicular to each other. It is possible that there were two large nuclei in the box after the initial heating from the seeding temperature and that these grow and clash into the other, stopping the other from filling the simulation cell. 

To further interrogate each of the different systems, the stem length distribution and structure factors are examined as seen in Figure \ref{fig:structure}. Stem length distributions are calculated by examining the local alignment of bond vectors using the local $P_2$ order parameter, commonly employed to assess the crystallinity in simulations of PE crystallisation \cite{fall2022role,hall2019divining,meyer2002formation}. The stem length distributions reflect the number of unbroken bonds in near straight chains where the bond angle lies approximately between 150$^{\circ}$ to 180$^{\circ}$. It highlights the contrast between systems seeded at different temperatures as well as between the different crystallisation methods. 

Figure \ref{fig:structure} (a) and (b) show the stem length distributions for the self-seeded and continuously cooled systems respectively. Note each of the curves are shifted on the y-axis in intervals of 0.02 \Will{for easier} comparison between the respective curves. It is immediately apparent that the distributions exhibit multiple pronounced peaks which correspond to integer multiples of the distance between neighbouring \Will{dibromo}s. This is indicated by the vertical dotted lines in each of the respective panels. In addition, the stem lengths appear to shift into longer multiples with increasing crystallisation temperature, see Fig \ref{fig:structure} (a), suggesting that the traditional lamellar thickening seen with increasing crystallisation temperature has become quantized. The system crystallised at $T_{\mathrm{c}}=67^\circ$C (red curve) exhibits two peaks corresponding to 2 and 3 bromine segments, $\sim$5 nm and $\sim$7.5 nm respectively, likely resulting from the two differently oriented domains with different thickness as shown in the snapshot in Fig. \ref{fig:Snaps} (a). Raising the crystallisation temperature to $T_{\mathrm{c}}=72^\circ$C results in a shift to longer stem lengths as shown by the blue curve, with the sharpest peak at 3 bromine segments ($\sim$7.5 nm) and a smaller less pronounced peak at 4 bromine segments ($\sim$11 nm). Whilst the stability of longer stems at higher crystallisation temperatures is not unexpected, it appears that this process is controlled by preferential folding at bromine groups along the chain. On raising the crystallisation temperature further still to $T_{\mathrm{c}}=77^\circ$C, most stems shift into the 4 ($\sim$11 nm) bromine segment regime, further confirming this observation. In comparison, the systems grown by continuous-cooling show very different distributions with weakly pronounced peaks only being observed with the slowest cooling rates ($<1.0$ K/ns) for 1 and 2 bromine segments at $\sim$2.5 nm and $\sim$5 nm respectively. The stem length distributions for these system systems drop quickly following a power law and the peaks are broader and shorter in comparison to the self-seeded systems. This suggests that very few stems can grow and that the majority are shorter and belong to less ordered crystal domains. It is worth noting, the quantized lamellar thickening seen here has not been witnessed in molecular simulations before. In linear PE chains, lamellar thickening usually appears gradual with a gaussian-like distribution of chain lengths \cite{fall2022role,meyer2002formation,fall2023molecular,hall2019divining}. 

This behaviour is further confirmed on inspection of the static structure factor, which is defined as
\begin{equation}
S(q) = \frac{1}{M_{tot}} \langle \sum_{i,j=1}^{M_{tot}} \exp{i.q(r_{i}-r_{j})} \rangle_{|q|=q \pm dq}
\label{e:sq}
\end{equation}
where the sum is performed overall monomers $M_{tot}=MN$ in the system (with M chains and N united monomers per chain) and the angular brackets indicate averaging overall q-vectors of length $q \pm dq$. A running average is applied up to the size of the box, 10.3nm. Because only q-vectors compatible with the finite box size may be considered, the precision becomes increasingly poor as inverse $q$ approaches the box size. The analysis was performed within the ranges of both the SAXS(low) and WAXS(high) q-values to get a comprehensive picture of large scale crystal features and the smaller atomic ones. Note the curves are shifted on the y-axis arbitrarily on a log scale \Will{for easier} comparison between systems. The WAXS range shows strong peaks which correspond to the hexagonal lattice harmonics. For the self-seeded systems, the first sharp peak $q_{0}$, is seen at q = 13.98 , 14.1 and 14.1 ${\mathrm{nm}}^{-1}$ for $T_{\mathrm{c}}$ = 67$^{\circ}$C, 72$^{\circ}$C and 77$^{\circ}$C, respectively \Will{corresponding to an interchain distance of 0.45 nm, as expected in crystalline PE and the hexagonal rotator phase of n-alkanes \cite{ungar1983structure}}. There are higher harmonics as well, which occur at q-values that are multiples of the first peak (at $\sqrt{3}$, 2, and $\sqrt{7}$ times the value of $q_{0}$). 

The hexagonal lattice is only weakly present in the continuously cooled systems with one diffuse peak for the 1.0 K/ns, 0.1 K/ns and the 0.04 K/ns cooling rates appearing at q = 13.7, 13.98, 14.14${\mathrm{nm}}^{-1}$ respectively. The rest of the peaks (harmonics) disappear because the crystalline structures are poorly ordered and there is no periodicity in the structure. The SAXS range is relatively noisy as inverse $q$ approaches the box size, however amidst the noise strong peaks corresponding to the different stem lengths are observed. This is more clear in the case of the self-seeded systems in panel (c), with lines corresponding to the stem lengths lining up perfectly with some of the peaks of the stem length distribution. The stem lengths are an average, which is why some of peaks do not align perfectly with the dotted lines corresponding to the length between successive \Will{dibromo}s along the chains. The peaks appear weakly for the slowest continuous-cooling rate in panel (d) and are entirely absent with faster cooling. 

\begin{figure}
    \centering
    \includegraphics[width=\columnwidth]{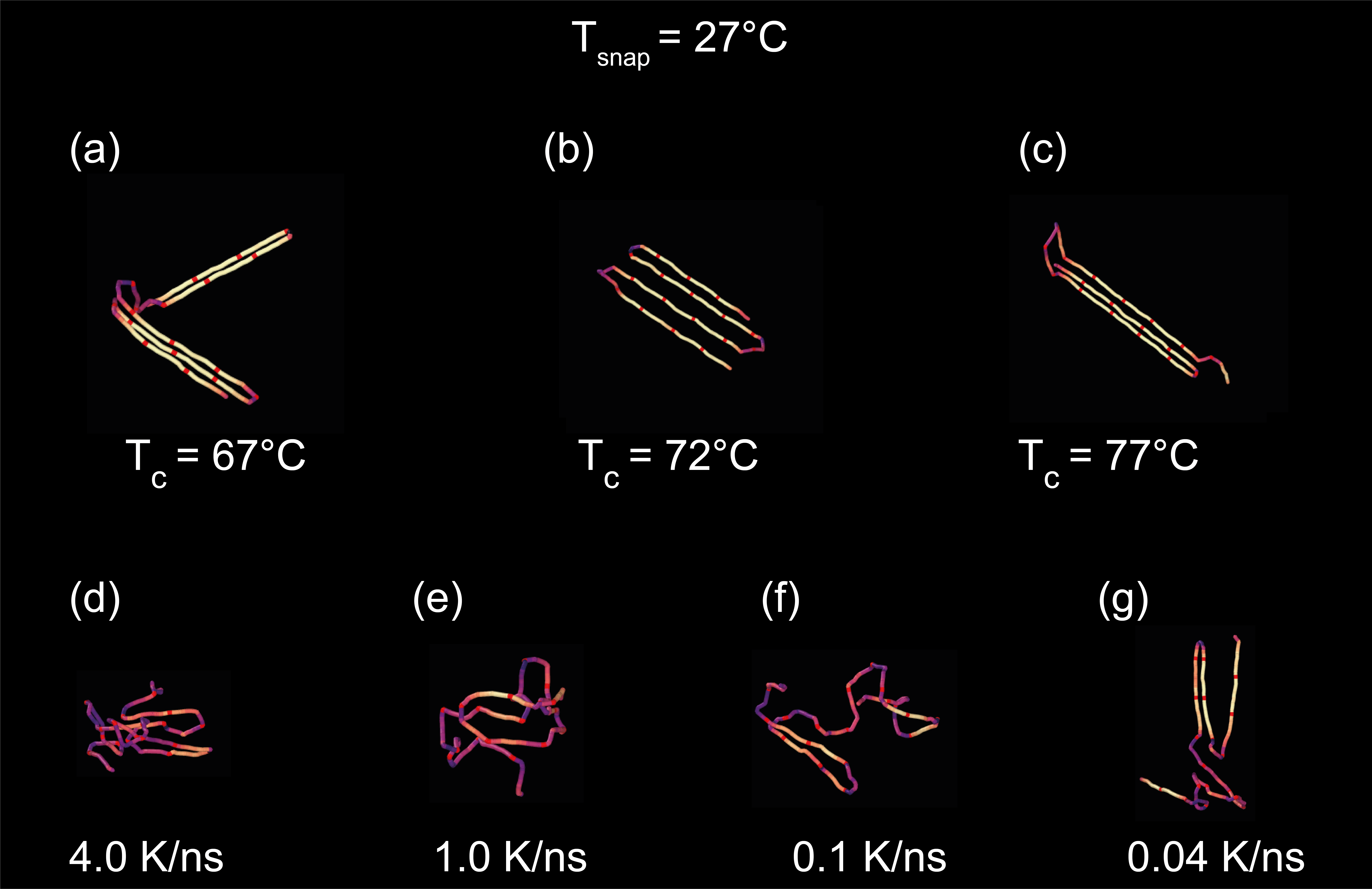}
    \caption{Snapshots of typical chain conformations in the self-seeded and continuously-cooled crystals at room temperature, 27$^{\circ}$C. Panels (a-c) correspond to systems crystallised at 67$^{\circ}$C, 72$^{\circ}$C and 77$^{\circ}$C and panels (d-f) correspond to cooling rates 1.0K/ns, 0.1K/ns and 0.04K/ns respectively. Polymer chains are coloured continuously according to their local $P_{2}$ order parameter from $P_{2}=0$ (purple, amorphous) to $P_{2}=1$ (yellow, crystalline) and the bromine species is shown in red. Note the pronounced folding at \Will{dibromo} groups and increasing stem lengths with crystallisation temperature in panels (a-c) not present in the continuously cooled systems.}
    \label{fig:stem}
\end{figure}

\begin{figure*}
    \centering
    \includegraphics[width=1.5\columnwidth]{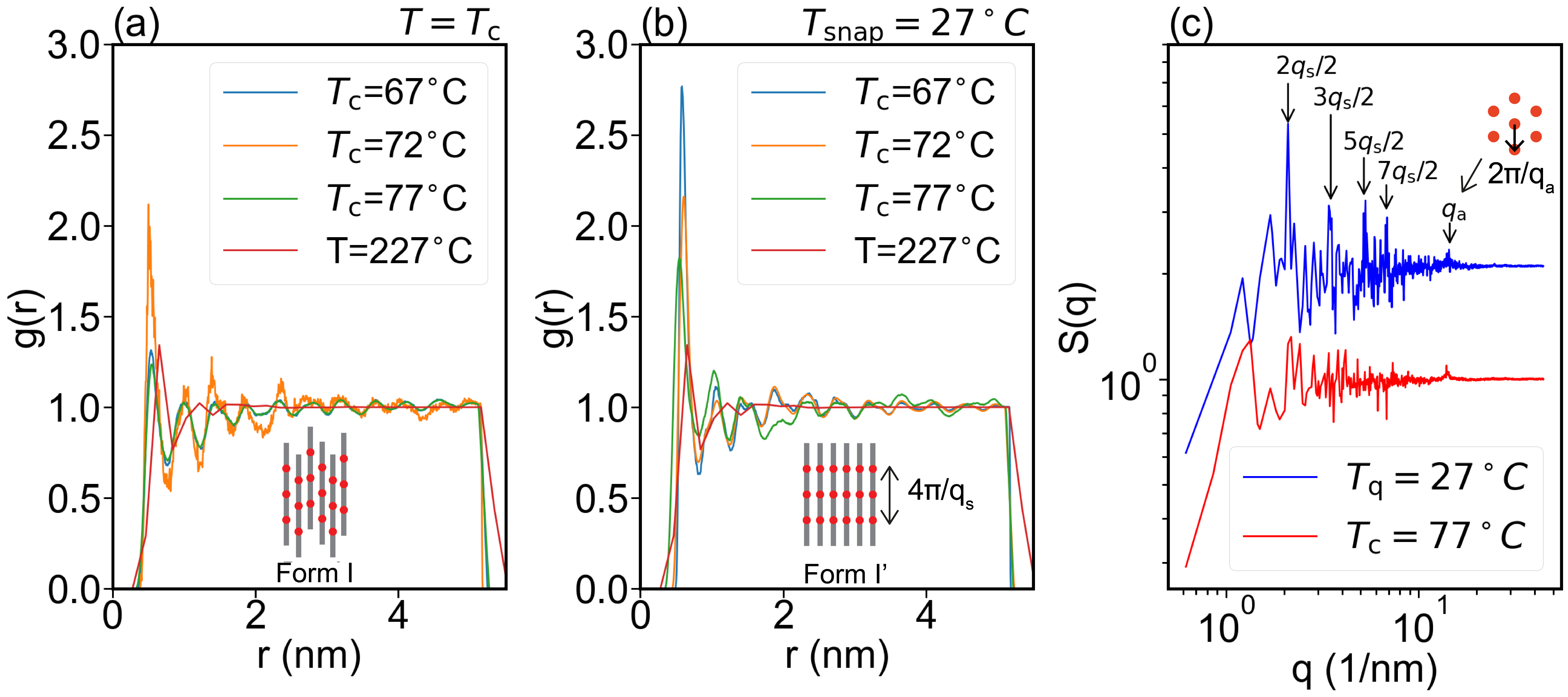}
    \caption{The radial distribution function calculated using only united-monomers containing bromine groups at the crystallization temperature and quench temperature and the structure factor for the bromines in the system crystallised at 77$^{\circ}$C, before and after quenching. (a-b) RDF calculated with 1200 bins, with a cutoff distance of 7.5nm, for the self-seeded systems at $T=T_{\mathrm{c}}$ and $T=27^{\circ}$C respectively, the RDF of the melt at 227$^{\circ}$C is shown in both panels for comparison. In panel (c) the partial structure factor is calculated using only the united-monomers containing bromine groups at the crystallisation temperature (red curve) and room temperature (blue curve).}
    \label{fig:gofr}
\end{figure*}

The controlled folding reflected in the stem length distributions and structure factors in Fig \ref{fig:structure} can be further revealed on inspection of the individual chain conformations within the crystalline domains. Representative examples for all systems are shown in Fig \ref{fig:stem}, where self-seeded and continuously cooled systems correspond to panels (a-c) and (d-g) respectively. In panel (a) at $T_{\mathrm{c}}$ = 67$^{\circ}$C a single 200 monomer long chain is shown spanning both crystal domains in the box. Interestingly clear tight folding at the (red) bromine groups can be seen confirming its function as a preferred fold site with 4 \Will{dibromo} groups per stem. At higher crystallisation temperatures similar folding is also seen but with longer stems of 5 and 6 bromine groups per stem at $T_{\mathrm{c}}$ = 72$^{\circ}$C and 77$^{\circ}$C respectively. Some apparent registry between neighbouring bromine units is also apparent which is further analysed in the \Will{preceding} section. Interestingly, the chain conformations in the continuously cooled systems in panels (d-f) also show a strong tendency to fold at the \Will{dibromo} groups but the stems do not fully crystallise.

To quantify the extent of registry between bromine units in the crystalline part of the lamella, the radial distribution function (RDF) between bromine species is calculated in Fig \ref{fig:gofr} (a) and (b) for self-seeded systems at $T=T_{\mathrm{c}}$ and $T=27^{\circ}$C respectively. Note, the crystallisation temperature or quench temperature are shown at the top right of the respective panels.  All self-seeded systems at the crystallisation temperature in panel (a) show a repeating series of maxima corresponding to neighbouring shells of bromine groups sitting next to one another in the crystal, which persist until the RDF reaches half the box size. Notably these peaks, whilst pronounced, are relatively broad and diffuse indicating the \Will{dibromo} units are not well registered. For reference the (red) curve indicates the RDF of the \Will{dibromo} units in the melt where no peaks are present beyond the first shell. 

In comparison, the RDF at room temperature in panel (b) shows sharper and more pronounced peaks which persist at large distances suggesting strong local registry is present. A partial structure factor using only bromine species is shown alongside in panel (c) for one representative system ($T_{\mathrm{c}}$ = 77$^{\circ}$C) both at the crystallisation temperature (red curve) and after quenching to room temperature (blue curve). Note at the crystallisation temperature only the hexagonal lattice of the rotator phase is present at $q_{\mathrm{a}}$, with no sharp Bragg peaks being observed at low $q$-values. However, on quenching, very sharp peaks appear at multiples of the layer spacing around $2q_{\mathrm{s}}/2\approx2.2$ (1 layer) with higher harmonics at $3q_{\mathrm{s}}/2$, $5q_{\mathrm{s}}/2$ and $7q_{\mathrm{s}}/2$. This corresponds to the layer spacing between bromine layers within the crystalline part of the lamella domain which are then tilted with respect to the layer normal. This is best illustrated in Fig \ref{fig:Snaps} (a-c) where pronounced registry can be seen between the layers not unlike that of a Smectic-C phase in liquid crystals. Due to thermal fluctuations it is challenging to reveal the ordering of the layers in a single snapshot and it is only after a coarse thermally averaged $S(q)$ using only the \Will{dibromo} species that the registered layers could be accurately quantified. The smaller peak at $q_{\mathrm{a}}\approx15$ corresponds to the hexagonal lattice parameter of neighbouring \Will{dibromo}s sitting side-by-side within each registered layer. 

Whilst isothermal temperature dependent lamellar thickening has been well known for several decades \cite{weeks1963melting,hoffman1965x}, the quantized behaviour shown here is rather rare. This \Will{phenomenon} has only been seen in a handful of macromolecular systems, including monodisperse ultra-long n-alkanes \cite{zeng2007semicrystalline,higgs1994growth,ungar1985crystallization,ungar2005effect} and more recently, polydisperse polyethylene chains with regularly placed chemical moieties or defects \cite{zhang2018effect,marxsen2022crystallization}, polyacetals \cite{zhang2019crystallization,zhang2020crystallization}, long-spaced aliphatic polyesters \cite{marxsen2020isothermal} and low molecular weight fractions of polyethylene oxide \cite{cheng1991nonintegral}. Several simulation studies of monodisperse polyethylene chains have demonstrated some form of stepwise lamellar thickening whether it be by simple folding \cite{hu2001chain,verho2018crystal} or by the introduction of short chain branching and controlled folding \cite{fall2023molecular,fall2022role}. However, none so far have been able to demonstrate quantized lamella thickening as we do here by the introduction of preferential fold sites along the chains.

\begin{figure}
    \centering
    \includegraphics[width=\columnwidth]{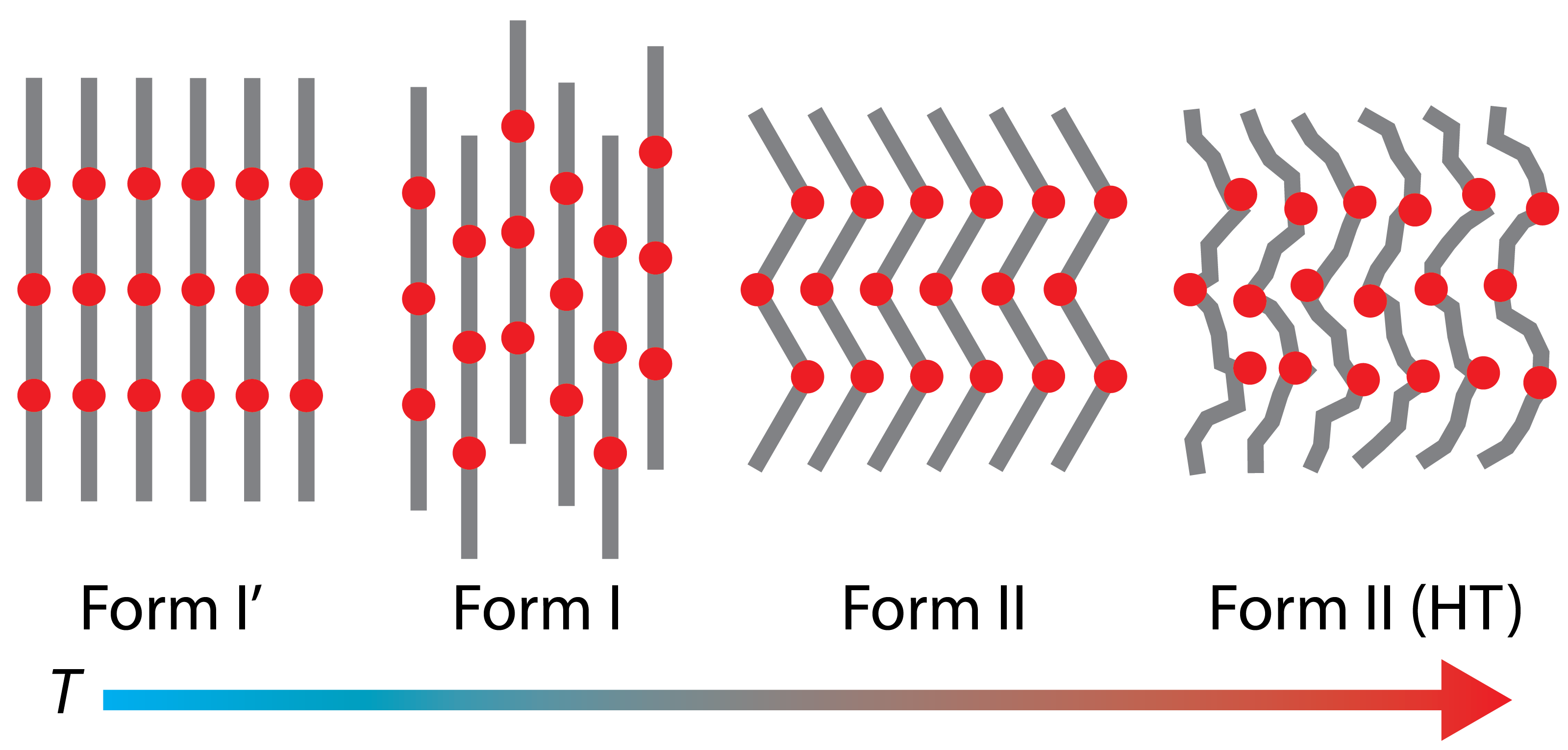}
    \caption{Cartoon polymorphs exhibited by precision PE chains with \Will{dibromo} defects placed regularly on every 21st carbon atom \cite{tasaki2014polymorphism}.}
    \label{fig:forms}
\end{figure}

The motivation for the introduction of regularly placed bromine defects stems from recent experimental studies by Tasaki et al \cite{tasaki2014polymorphism}. There it was found that the polymer had four crystalline phases or `polymorphs' called Form I, Form I${^\prime}$, Form II and HT, see Fig \ref{fig:forms}. In Form I and Form I' polymer chains are fully aligned in an all-trans conformation in the traditional textbook structure of a polymer crystal. Within the lamellae the bromines are incorporated fully into the crystal, not unlike short methyl or ethyl branches which are well-known to at least partially crystallise in polyethylene \cite{zhang2018direct,zeng2007semicrystalline}. At high temperature, in Form I, the bromines within the lamellae are partially disordered within the crystalline lamellar stacks. On quenching however, the bromines group together into registered layers with pronounced positional order, resembling a Smectic C phase not unlike that seen in liquid crystals made of much smaller molecules. This registered arrangement of Form I, is known as Form I${^\prime}$. The sudden onset of positional ordering of the bromine layers, seen on quenching from $T_{\mathrm{c}}$ to $T_{\mathrm{q}}$ in the self-seeded systems in Fig \ref{fig:structure}, suggests our simulations are indeed representative of real experiments despite the simplicity of the coarse-grained model employed. Likely this behaviour is brought on by the repulsion between neighbouring \Will{dibromo} units, as remarked in \cite{tasaki2014polymorphism}, which partially disrupts the layering at higher temperatures. Further high temperature forms are also observed experimentally, with a chevron type form resembling that of an anticlinic Smectic C phase, where bromines positionally order at the vertices of alternating chevron layers, see Fig \ref{fig:forms}. This further morphs into a high temperature (HT) form where registry between bromine groups is again partially disrupted. Due to the small box sizes used here ($\sim$ 10 nm), it is not possible to grow Form II or HT polymorphs in our simulations, work in this direction is currently underway in bigger systems.  

In all systems which exhibit quantized thickening, the appearance of self-poisoning also appears to be ubiquitous \cite{whitelam2016minimal,marxsen2022crystallization,zhang2018effect}. In this process, metastable polymorphs with different thicknesses compete at the growth front during crystallisation. \Will{Above} the melting point of a thinner form \Will{but close to it}, frequent attachments of metastable short stems can inhibit the growth of the thicker form poisoning its growth. Only once the shorter stems are removed can the thicker form continue to grow and this manifests \Will{itself} as a pronounced growth rate minimum in the crystallisation rate with temperature. This phenomenon has been a topic of debate for several decades since it was first discovered in \cite{ungar1987inversion} and theoretically analysed in \cite{higgs1994growth}\Will{. It} has been the subject of several theoretical and computational studies since \cite{hu2001chain,whitelam2016minimal,gabana2024quantitative,higgs1994growth,ma2009understanding}. Whilst we do observe the coexistence of different lamellar thicknesses, the 1d picture of self-poisoning is not directly witnessed here, nonetheless our results do indeed support a complex 3d and growth front driven view of polymer crystallisation.

\section{Conclusions}

By using the united-monomer model of PE and extending it to include \Will{dibromo} groups, it became possible to access semi-crystalline structures on length-scales typical of PE crystals. In addition, a careful choice of crystallisation protocol, known as self-seeding, enabled the growth of precision PE crystals. This allowed us to demonstrate quantized lamellar thickening, of PE chains with regularly spaced brominated groups, with increasing crystallisation temperature and how this is driven by regular placement of \Will{dibromo} \Will{units} along the PE backbone. The observed behaviour, was almost completely obscured using the traditional continuous-cooling protocol, which is most commonly employed in simulations of PE. A regular placement of Br groups along the backbone, appears to limit the stem length not through rejection of Br groups from the crystal but instead by \Will{preferential} folding at these groups. In combination with temperature dependent lamellar thickening, this provides a new route to \Will{controlling} the lamellar thickness, in integer multiples of the spacing between successive Br groups. At the crystallisation temperature, the Br groups appeared well tolerated within the crystal however quenching to room temperature revealed the tendency of Br groups to form registered layers within the crystalline part of the lamellae.  

These systems now present an opportunity to understand how changes in molecular architecture such as the regular placement of small chemical moieties, halogens or short chain branches, influence the semi-crystalline morphology in great detail. Building on the results presented here, it will be interesting to investigate how the spacing and number of Br groups influences the semi-crystalline morphology and controls the lamellar thickening. In addition, larger systems may be considered to investigate recently reported chevron shaped crystals, with lamellae some 70nm in thickness, and whether the united-monomer model can faithfully reproduce the essential features. Work in these directions is currently underway. 






\section*{Acknowledgements}
 We thank the High Performance Computing Center CAIUS of the University of Strasbourg for supporting this work by providing access to computing resources, partly funded by the Equipex Equip@Meso project (Programme Investissements d'Avenir) and the CPER Alsacalcul/Big Data. The ShARC (Sheffield Advanced Research Computer), one of the University of Sheffield’s High Performance Computing Clusters is gratefully acknowledged. A generous grant on the ARCHER2 UK National Supercomputing Service (https://www.archer2.ac.uk) is also gratefully acknowledged. \Will{G.U. gratefully acknowledges support from National Science Foundation of China (22250710137).} X.Z. thanks support from the UK Engineering and Physical Sciences Research Council (EP-T003294).


\bibliography{gabana_et_al}

\section*{Supplementary info}

\section{All-Atom Simulations}

Similarly to our previous work on united-monomer models i.e. CG-PVA \cite{meyer2001formation} or CG-PE \cite{fall2022role}, the coarse-graining scheme utlilised respresents one chemical monomer (C$_{2}$H$_{4}$) as a single coarse-grained bead in the simulation. Non-bonded Lennard-Jones interactions and angular potentials are obtained via Boltzmann inversion of the distribution functions extracted from all-atom simulations, using every second backbone carbon as distribution centres \cite{vettorel2006coarse}. Choosing this mapping ensures cross-correlations between angles or bonds are small \cite{reith2001mapping} and the angular states of the CG model map directly to all-atomistic conformers, see Fig. 1. in the main manuscript.

The CG potentials of the CG-PE \cite{fall2022role} model remain unchanged since bonds and radial distribution functions are similar however including bromide dimers (CH$_{2}$CBr$_{2}$) required new all-atom simulations. In keeping with CG-PE, all atom simulations use the OPLS force field \cite{JoMaTR96jacs} with a modified torsional potential. It is well known that OPLS-AA overestimates the crystallisation temperature \cite{paraffinreview2019rsc} and this is due to the gauche energy of the backbone torsions being too high thus favouring the extended state. We perform simulations of C$_{10}$H$_{21}$CBr$_{2}$C$_{9}$H$_{19}$ where the leading torsional term has been lowered from 1.76 to 0.8 kcal/mol. All simulations were performed identically to the CG-PE model at 500K sufficiently far above the melting temperature.  Where the bromide unit is concerned only the size of the Br units and C-Br bond lengths are modified, using the parameters for halogens found in \cite{jorgensen2012treatment}, the modified torsional potential of OPLS-AA is utilised throughout the entire backbone. 

\section{Coarse-Grained Model Development}
\newpage

\begin{figure*}
    \centering
    \includegraphics[width=1.8\columnwidth]{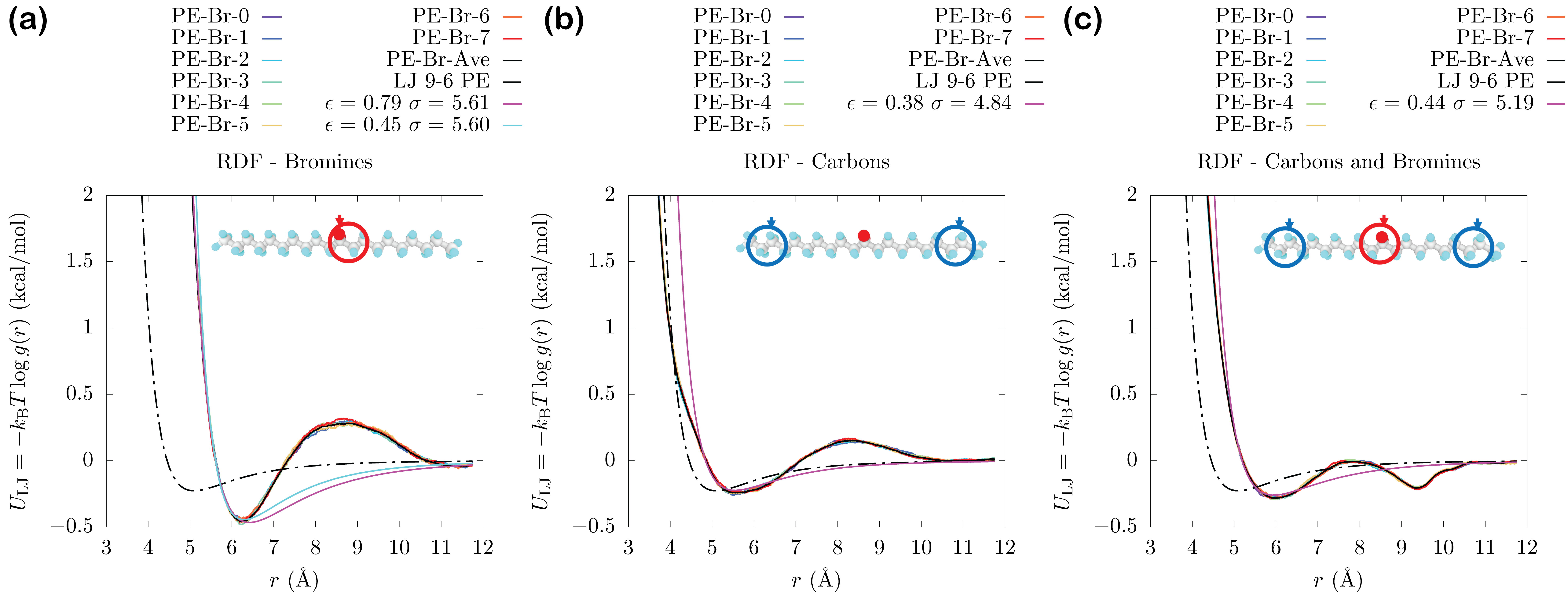}
    \caption{Radial distribution functions (RDFs) obtained from all-atom simulations of 192  C$_{10}$H$_{21}$CBr$_{2}$C$_{9}$H$_{19}$ chains in the melt at 500K between coarse-grained centers. (a) Bromide-Bromide, (b) Carbon-Carbon and (c) Bromide-Carbon. The inset cartoon chains, illustrate the positions of the CG centers used when calculating the RDF, blue and red circles represent C$_{2}$H$_{4}$ and CH$_{2}$CBr$_{2}$ units respectively. The black line represents the average over all simulation packages, purple and cyan lines correspond to the best fitting of 9-6 or 12-6 LJ potentials respectively. Dashed lines represent the CG-PE model. Note the strong overlap of the potentials obtained from AA simulation with the original CG-PE model in panel (b), thus we choose to use the CG parameters from the CG-PE model to describe the interaction between standard PE units.}
    \label{fig:lj_potentials}
\end{figure*}

\begin{figure*}
    \centering
    \includegraphics[width=1.8\columnwidth]{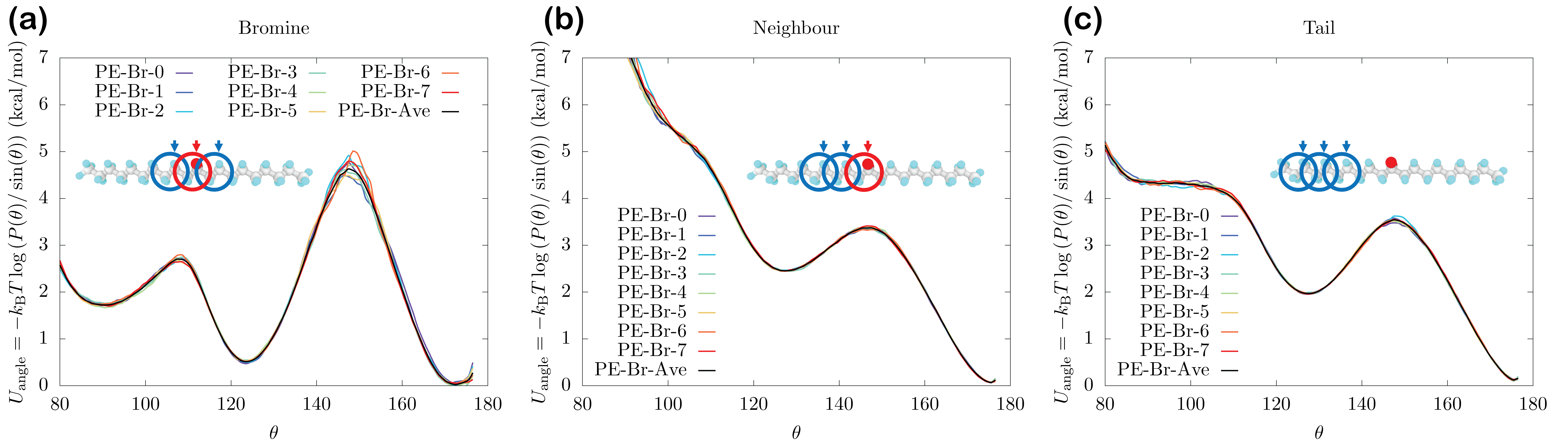}
    \caption{Angular potentials obtained from all-atom simulations of 192  C$_{10}$H$_{21}$CBr$_{2}$C$_{9}$H$_{19}$ chains in the melt at 500K between coarse-grained centers. The three panels correspond to CG angles with (a) the Bromide unit at its centre, (b) the bromide unit at its tail and (c) between only CG units containing carbons. The inset cartoon chains, illustrate the positions of the CG centers used when calculating the potentials, blue and red circles represent C$_{2}$H$_{4}$ and CH$_{2}$CBr$_{2}$ units respectively. At the bromide unit a pronounced \textit{gauche-gauche} minimum appears with a higher torsional barrier separating the \textit{trans-gauche} and \textit{trans-trans} stares. Close to the bromide unit the \textit{gauche-gauche} minimum appears completely supressed. Note the potentials obtained between standard PE units (the tails) in panel (c) are near identical to the standard PE model, thus remain unchanged.}
    \label{fig:ang_potentials}
\end{figure*}

\end{document}